\documentclass[%
 reprint,
nobibnotes,
 amsmath,amssymb,
 aps,
 prl,
]{revtex4-1}

\usepackage{graphicx}
\usepackage{dcolumn}
\usepackage{bm}

\usepackage{xcolor}

\begin{document}

\title{Machine learnt approximations to the bridge function \\ yield improved closures for the Ornstein-Zernike equation.}

\author{Rhys E. A. Goodall}
    \affiliation{Cavendish Laboratory, University of Cambridge, Cambridge, UK}
    
\author{Alpha A. Lee}
    \email[Correspondence email address: ]{aal44@cam.ac.uk\\}
    \affiliation{Cavendish Laboratory, University of Cambridge, Cambridge, UK}

\begin{abstract}
A key challenge for soft materials design and coarse-graining simulations is determining interaction potentials between components that give rise to desired condensed-phase structures. In theory, the Ornstein-Zernike equation provides an elegant framework for solving this inverse problem. Pioneering work in liquid state theory derived analytical closures for the framework. However, these analytical closures are approximations, valid only for specific classes of interaction potentials. In this work, we combine the physics of liquid state theory with machine learning to infer a closure directly from simulation data. The resulting closure is more accurate than commonly used closures across a broad range of interaction potentials. We show for \textcolor{black}{two examples of} a prototypical inverse design problem, fitting a coarse-grained simulation potential, that our approach leads to improved one-step inversion.
\end{abstract}


\maketitle

\section{Introduction}

A central question in soft matter pertains to the inverse problem of determining the interaction potentials between building blocks, e.g. colloids or molecules, that give rise to desired structures through self-assembly \cite{jadrich2017probabilistic,sherman2020inverse}. Applications of this inverse problem abound in disparate fields. For instance, molecular interactions can be optimised to yield porous structures in liquids \cite{giri2015liquids} which in turn are crucial for chemical processes such as gas separation and storage \cite{zhang2015porous, zhang2016thermodynamics}. \textcolor{black}{Studying the inverse problem for these porous structures, both on computationally designed systems \cite{lindquist2016assembly} and by back-tracking potentials from experimental data, allows for greater understanding of their physics which can be leveraged to further optimise these systems and improve their properties.} Similarly, \textcolor{black}{bottom-up} coarse-graining \cite{kmiecik2016coarse,marrink2007martini,brini2013systematic}, an approach for accelerating soft matter simulations, \textcolor{black}{can be viewed as an inverse problem. Coarse-graining }involves finding effective interparticle interactions between coarse-grained ``beads'' that reproduce the structure of the full system. 

\textcolor{black}{The inverse problem we consider in this work is given the pair distribution function $g(r)$ of an isotropic collection of particles with density $\rho$ can we determine a pairwise interaction potential $\phi(r)$ that would result in the original pair distribution $g(r)$ being recovered from a forward simulation of particles with the same density.} Although the forward problem of predicting condensed phase structure given a set of interactions can be tackled with standard methods such as molecular dynamics or Monte Carlo simulation, solving the inverse problem remains challenging. \textcolor{black}{Iterative Boltzmann Inversion (IBI) \cite{schommers1983pair} and its extensions \cite{reith2003deriving, moore2014derivation} are perhaps the most common approaches. These methods involve iterative optimisation loops where the trial potential is updated based on the differences between the target $g(r)$ and that obtained from converged simulations of the trial potential. Model-free predictor-correction methods \cite{stones2019model} solve a similar iterative problem to IBI but do so via using test-particle sampling to obtain estimates for the $g(r)$ resulting from a given trial potential, therefore, avoiding the need for multiple sets of simulations to be carried out. Other approaches include direct inversion schemes such as the generalized Yvon-Born-Green method \cite{mullinax2009generalized, rudzinski2015generalized} that approximates the potential of mean-force from structural correlation functions.} 

In theory, a rigorous framework in statistical physics known as the Ornstein-Zernike equation \cite{ornstein1914accidental} provides a direct and computationally efficient framework to solve this inverse problem without iterative approaches.  In an isotropic fluid with density $\rho$ and pair distribution function $g(r)$, the Ornstein-Zernike equation defines the direct correlation function $c(r)$ in terms of total correlation function, $h(r)=g(r)-1$, via
\begin{equation}\label{eq:OZ}
h(r) = c(r)+ \rho\int c(|r-r'|)h(r') dr'.
\end{equation} 
The key insight is that the total correlation function is a consequence of not only direct interactions between particles but also indirect correlations mediated through interactions with other particles.

Given a closure relationship coupling $h(r)$ and $c(r)$ with the interaction potential, $\phi(r)$, the Ornstein-Zernike equation provides a path to solve the inverse problem. The generally accepted form for the closure is 
\begin{equation}\label{eq:closure}
h(r) + 1 = \exp\big(-\beta\phi(r) + \gamma(r) + B(r)\big),
\end{equation}
where $B(r)$ is the bridge function and $\gamma(r) = h(r)-c(r)$ is the indirect correlation function. Whilst diagrammatic expansions exist that define $B(r)$, for practical applications no convenient closed-form solution exists \cite{hansen2006theory}. As a result, $B(r)$ has traditionally been approximated with a functional in terms of $\gamma(r)$. The most common are the Hyper-netted Chain approximation (HNC) \cite{HNC1958closure}, $B^{HNC}(r) = 0$, and the Percus-Yevick approximation (PY) \cite{PY1958closure}, $B^{PY}(r) = \ln \big(1+\gamma(r) \big) - \gamma(r)$. HNC is well suited for long-range potentials, whilst PY provides an analytical solution for the hard-sphere case \cite{Wertheim1963pyhard} and works well for short-range, purely repulsive systems. 

Modern machine learning (ML) offers a suite of powerful tools for function approximation \cite{breiman2001random,williams2006gaussian,schmidt09eureqa,goodfellow2016deep} that have accordingly attracted attention in many areas within the physical sciences \cite{butler2018machine, carleo2019ml4phy}. In this work, we show that closures to the Ornstein-Zernike framework can be learnt directly from simulation data by approximating $B(r)$ using an ML model. The results indicate that when used to solve the inverse problem such learnt closures yield better estimates of the bridge function and therefore the potential than either HNC or PY across a broad range of systems. We apply a learnt closure in \textcolor{black}{two examples of} a prototypical coarse-graining problem and show that it yields improved one-step inversion.  

\section{Methodology}

\subsection{Feature Set Design}

\textcolor{black}{Although the aim when solving the inverse problem is to determine an interaction potential $\phi(r)$ it is more convenient to consider approximating the bridge function $B(r)$ in \eqref{eq:closure} as opposed to the entire closure. Doing so allows our approach to be viewed as learning a correction to the successful HNC closure therefore building in a strong physical prior. This approach also presents a numerically easier learning problem because $B(r)$ does not diverge in regions where $\phi(r)$ diverges (this can be seen for the Lennard-Jones 6-12 potential in \cite{llano1992bridge} and for the Soft-Sphere potential in \cite{llano1994bridge}).} 

In general, the performance of an ML model is circumscribed by whether the information captured in the model's inputs is sufficient to determine the system. Therefore, physically motivated input features are necessary to learn $B(r)$ effectively. $B(r)$ can be expanded as an infinite series in $\gamma(r)$\cite{lee2008molecular},
\begin{equation}
\label{eq:bridgeF}
B(r) = \frac{\bar{F_3}}{2!} \gamma^2(r) + \frac{\bar{F_4}}{3!} \gamma^3(r) + ... ,
\end{equation}
where the average modification functions, $\bar{F_n}$, are dependent on the density, $\rho$, and the temperature, $T$. By dimensional analysis, the bridge function can only be expressed in terms of dimensionless reduced quantities $\rho^*$ and $T^*$. However, for complicated pairwise potentials, where multiple length and energy scales are required to define the system, comparable reduced quantities are ill-defined. This prevents the formulation of a truly general closure in terms of $\gamma(r)$ only. However, this also suggests that there is scope to improve upon current closures by including additional input features that allow us to recover the degrees of freedom corresponding to $\rho^*$ and $T^*$. This idea is apparent in how the modified Verlet closure \cite{choudhury2003integral} achieves significant performance improvements over the standard Verlet closure \textcolor{black}{(VM) \cite{verlet1980integral}, $B^{VM}(r) = - \gamma(r)^2 / 2 \big( 1+ \alpha \gamma(r) \big)$,} in systems where $\rho^*$ and $T^*$ are well defined by expressing the $\alpha$ parameter of the Verlet closure as a function of $\rho^*$ and $T^*$ directly. If $\rho^*$ and $T^*$ are not well defined similar increases in accuracy can be achieved by introducing additional parameters into the closure that can be fitted for known systems to ensure the self-consistency of thermodynamic properties. Examples of such closures are the Rogers-Young and Zerah-Hansen closures \cite{rogers1984closure,hansen1986closure} which introduce switching functions parameterised with characteristic length-scales that must be fitted. However, for a closure to be generally applicable any additional input features or parameters must be determined without prior knowledge of the target system.

From Equation \ref{eq:OZ}, we see that the density of the system can be extracted if both $h(r)$ and $c(r)$ are given. This suggests that taking $h(r)$ and $c(r)$ together as input features should be more informative than $\gamma(r)$ alone \cite{tsednee2019closure}. The liquid systems of interest exist in equilibria defined by detailed balance. Therefore, an additional feature can be identified by drawing inspiration from the fluctuation-dissipation theorem,
\begin{equation}
    \chi(r) = \frac{\langle(g(r)^2\rangle - \langle(g(r)\rangle^2}{\langle g(r)\rangle} \times \sqrt{N}, 
\end{equation}
where we scale by $\sqrt{N}$ to remove the dependence on the number of particles under observation, $N$. 

The final extension to the feature set considered was to include gradient information. This is done via $\gamma'(r)$ rather than $h'(r)$ and $c'(r)$ as the latter contain sharp jumps around the first co-ordination shell that cancel in $\gamma'(r)$. Such behaviour is undesirable as it would provide artefacts to which the model could over-fit on in the training data, leading to poor generalisation performance in downstream applications. In general, a desirable closure should be scale-invariant, however, the definition of a gradient requires a length-scale. To deal with this the radius of the first co-ordination shell is used as the reference length-scale in all systems. This allows gradient features to be defined in a consistent manner across systems. 

\subsection{\textcolor{black}{Model fitting and Data Generation}}

\begin{figure}
\centering
\includegraphics[width=3in]{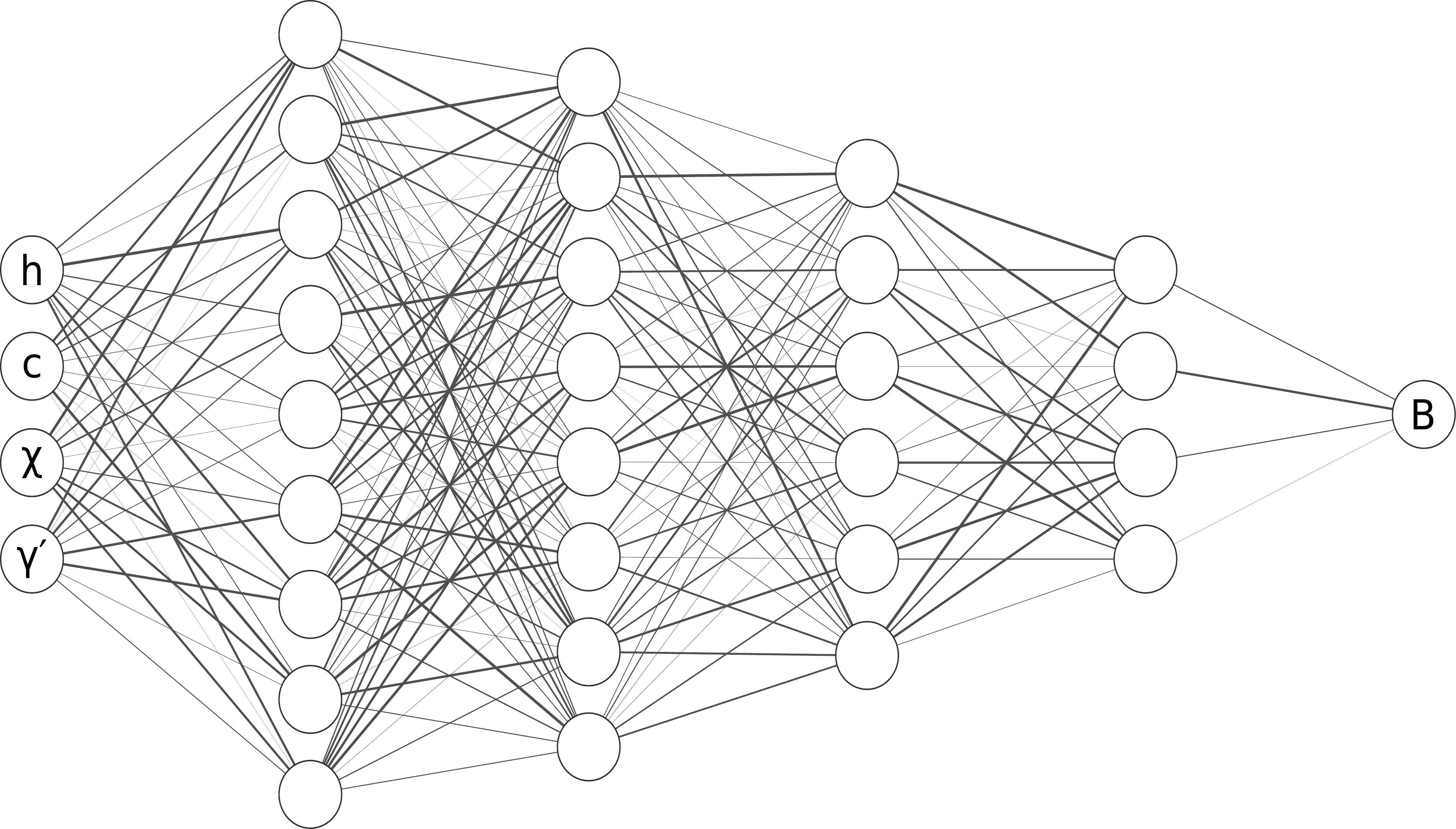}
\caption{\label{fig:nn}\textcolor{black}{\textbf{Schematic diagram of a Multi-Layer Perceptron.} The schematic shows a 5-layer multi-layer perceptron as used in this work. The input units take in $h(r)$, $c(r)$, $\chi(r)$ and $\gamma'(r)$ which are then transformed by a series of successive non-linear compositions to return a prediction for $B(r)$.}} 
\end{figure}

\textcolor{black}{In this problem data can be readily generated by solving the forward problem for specified potentials. When $\phi(r)$ is specified \eqref{eq:closure} can be inverted to calculate $B(r)$ if $h(r)$ and the  structure factor, $S(q)$, are measured from a converged forward simulation (measurements of $S(q)$ are needed to obtain more reliable estimates of $c(r)$ than possible from $h(r)$ alone due to finite size effects - full details in the Supplementary Materials).}

Having constructed a physically motivated feature set, we employ neural networks to learn \textcolor{black}{approximations for $B(r)$ as a functional of the identified input features $h(r)$, $c(r)$, $\chi(r)$, $\gamma'(r)$. To do this we make use of a class of neural networks called multi-layer perceptrons (MLP). A $l$-layer MLP approximates functions $f(x)$ by $l$ successive non-linear compositions, i.e. $f(x) \simeq W_l\sigma(W_{l-1}...\sigma(W_1x)))$, where $x$ is a vector of input features, $W_i \in \mathbb{R} ^{M_i \times M_{i-1}}$ is a weight matrix inferred from data, $M_i$ is the number of units in layer $i$, and $\sigma(x)$ is a non-linear activation function - here we use rectified linear unit (ReLU) activations of the form $ReLU(x)=\text{max}(x,0)$. We use a 5-layer MLP with $M = [256, 128, 64, 32]$ units in the 4 hidden layers and a single unit in the output layer (Schematic shown in Fig. \ref{fig:nn}). In neural networks the gradients of the loss with respect to the model parameters $W$ can be efficiently inferred via back-propagation \cite{bishop1995neural, goodfellow2016deep}. In this work the first-order stochastic optimiser Adam \cite{kingma2014adam} was used to fit or ``train'' the model by adjusting its parameters $W$ to minimise a squared error loss function between the model's predictions and training labels for $B(r)$ obtained from the simulation data. The models were trained for 100 epochs (cycles through the training data) in mini-batches of 128 samples at a time. The Adam optimiser was configured with a learning rate of 0.001, $\beta_1$ of 0.9, $\beta_2$ of 0.999, and $\epsilon$ of $10^{-7}$.}

\textcolor{black}{We use an MLP in this work as they are a numerically tractable way of representing complex functions that scale well to large data sets. However, there are no specific inductive biases built into the MLP and therefore alternative regression models \cite{breiman2001random,williams2006gaussian,schmidt09eureqa} would likely achieve similar performance. An example of a useful inductive bias is that the translational invariance of features/objects within images is automatically achieved in computer vision applications when using a convolutional neural network.}

Despite remarkable successes, neural networks are essentially powerful interpolation frameworks. Therefore, to learn a generally applicable closure, a wide variety of possible interaction potentials need to be explored when fitting the model. We investigate 13 different interaction potentials that we group into four classes: 
\begin{enumerate}
\item \textcolor{black}{Fast-Diverging} -- potentials containing strong divergences that prevent particles from overlapping,
\item \textcolor{black}{Step-Diverging} -- \textcolor{black}{Fast-Diverging} models where a repulsive plateau is added before the divergence to introduce complex multi-lengthscale structure, 
\item \textcolor{black}{Slow-Diverging} -- weakly divergent systems analogous to \textcolor{black}{Fast-Diverging} systems, and 
\item \textcolor{black}{Core-Overlapping} -- potentials that do not diverge and allow particles to overlap.
\end{enumerate} 
We will refer to \textcolor{black}{Fast-Diverging} and \textcolor{black}{Step-Diverging} potentials as hard potentials and \textcolor{black}{Slow-Diverging} and \textcolor{black}{Core-Overlapping} potentials as soft potentials. \textcolor{black}{In total we consider 96 different parameterisations of these 13 potentials by adjusting the length and energy scales}. For each of these \textcolor{black}{parameterisations} the molecular dynamics package \textit{ESPResSo} \cite{limbach06a, arnold13a} was used to determine $h(r)$ and $S(q)$ for systems of particles \textcolor{black}{with particle densities at increments of 0.1 between 0.4 and 0.8 giving a total of 480 systems}. Full details of the functional forms for the potentials investigated and simulation setup are available in the Supplementary Materials. 

\textcolor{black}{The Ornstein-Zernike formalism is only valid for fluid systems. Given that the high-throughput approach used to generate training data can result in simulations of non-fluid systems being carried out accidentally it is necessary to identify and exclude solid and two-phase samples that arise before the data can be used. This is done using several physically motivated heuristics. The Hansen-Verlet criterion \cite{march2002introduction}, $S(q_{peak}) > 2.8$, is used to identify solid and two-phase solid-liquid samples. Two-phase liquid-gas systems are identified using the heuristic criteria that $S(0) > 1$. This criterion is derived by noting that the compressibility of a system is given by:}
\begin{equation}\label{eq:compress}
\textcolor{black}{S(q \to 0) = \rho k_b T \kappa_T}
\end{equation}
\textcolor{black}{As gases are typically characterised by their highly compressible nature and noting that for an uncorrelated fluid $S(0) \simeq 1$ we propose that divergence of $S(q)$ in the limit $q \to 0$ is indicative of two-phase liquid-gas behaviour. We also check the consistency of the kinetic temperature across the simulation. In total 450 out of 480 systems investigated passed these heuristic criteria.}

\begin{table}
\centering
\caption{The coefficient of determination ($R^2$), Root Mean Squared Error (RMSE) and Mean Absolute Error (MAE) for different closures on a randomly held-out test set comprising 20\% of the simulation data.}
\label{tab:compare}
\begin{tabular}{lrrr}
\\
\hline
Closure   & \multicolumn{1}{c}{$R^2$} & \multicolumn{1}{c}{RMSE} & \multicolumn{1}{c}{MAE} \\
\hline
HNC                                  & 0.000 & 0.097 & 0.028 \\ 
PY                                  & 0.060 & 0.541 & 0.113 \\
VM ($\alpha$ = 0.8)               & 0.081 & 0.415 & 0.090 \\
G = $B(\gamma; r)$                 & -1.699 & 0.088 & 0.023 \\
HC = $B(h, c; r)$                   & 0.374 & 0.071 & 0.019 \\
HCX = $B(h, c, \chi; r)$             & 0.556 & 0.067 & 0.020 \\
LC = $B(h, c, \chi, \gamma'; r)$   & 0.693 & 0.052 & 0.017 \\
\hline
\end{tabular}
\end{table}

\begin{figure*}
    \centering
    \includegraphics[width=0.49\textwidth]{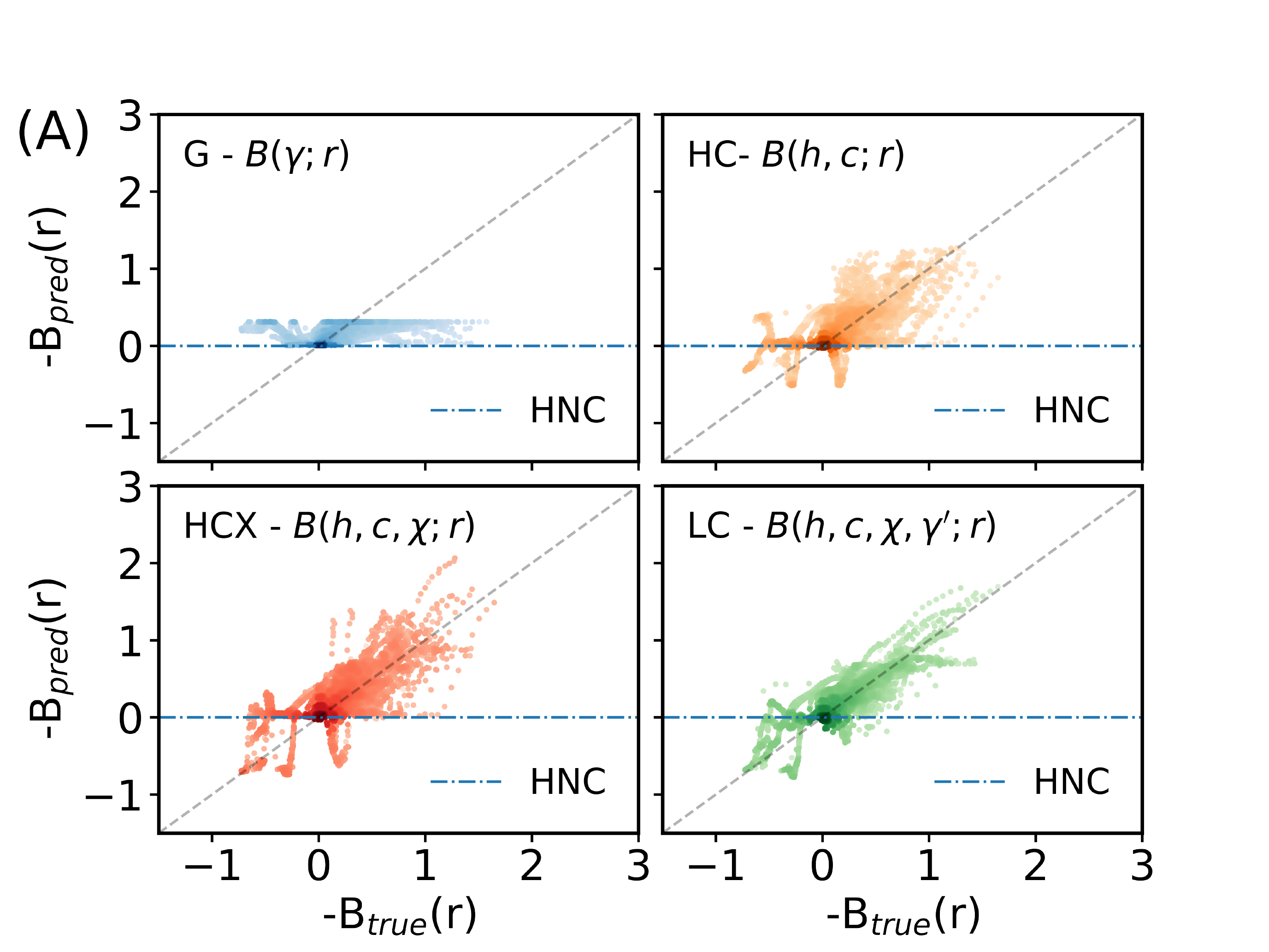}
    \includegraphics[width=0.49\textwidth]{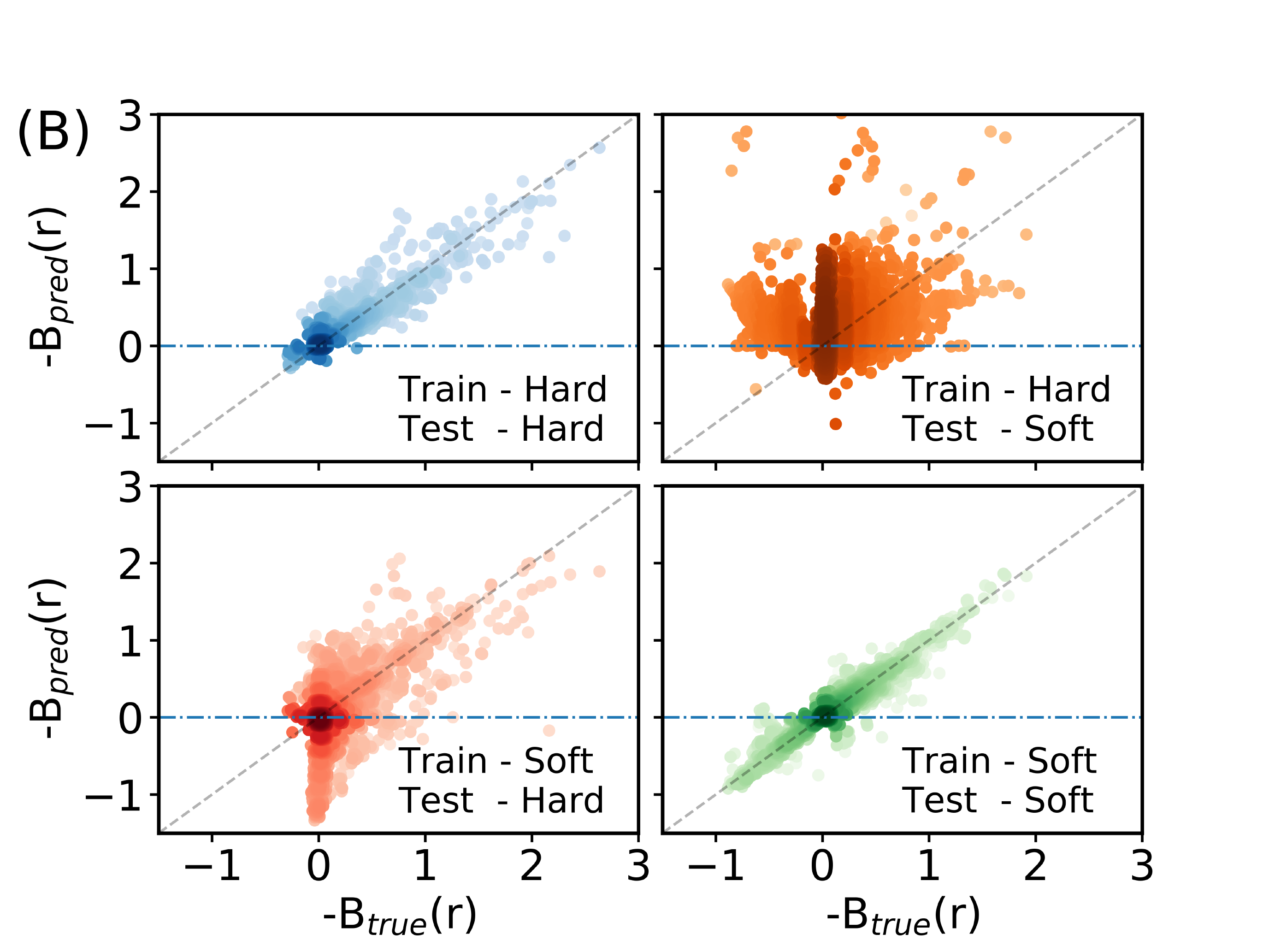}

    \caption{\textbf{Prediction-truth parity plots for learnt closures on held-out test sets.} \textcolor{black}{The plots show the prediction from a trained ML model on the y-axis against the true value on the x-axis for the test set data that was held-back during the model's training. An ideal model would result in the data sitting on a diagonal line $y=x$ as shown in grey. A blue-dashed line indicates results using the HNC closure $B^{HNC}(r)=0$}. The plots are shaded according to the log-density of points. (A) shows the performance of closures trained using different feature sets. As the feature set is extended the closures get better at predicting the value of bridge function, $B(r)$. The dark spots at the origin correspond to having learnt the correct far-field behaviour. (B) shows the performance of closures trained on restricted classes of potentials. Whilst the learnt closures are highly predictive when tested on potentials similar to those used to train them, they are less predictive in their out-of-training-distribution regimes.}
    \label{fig:parity-random}
\end{figure*}

\begin{figure*}
    \centering
    \includegraphics[width=0.49\textwidth]{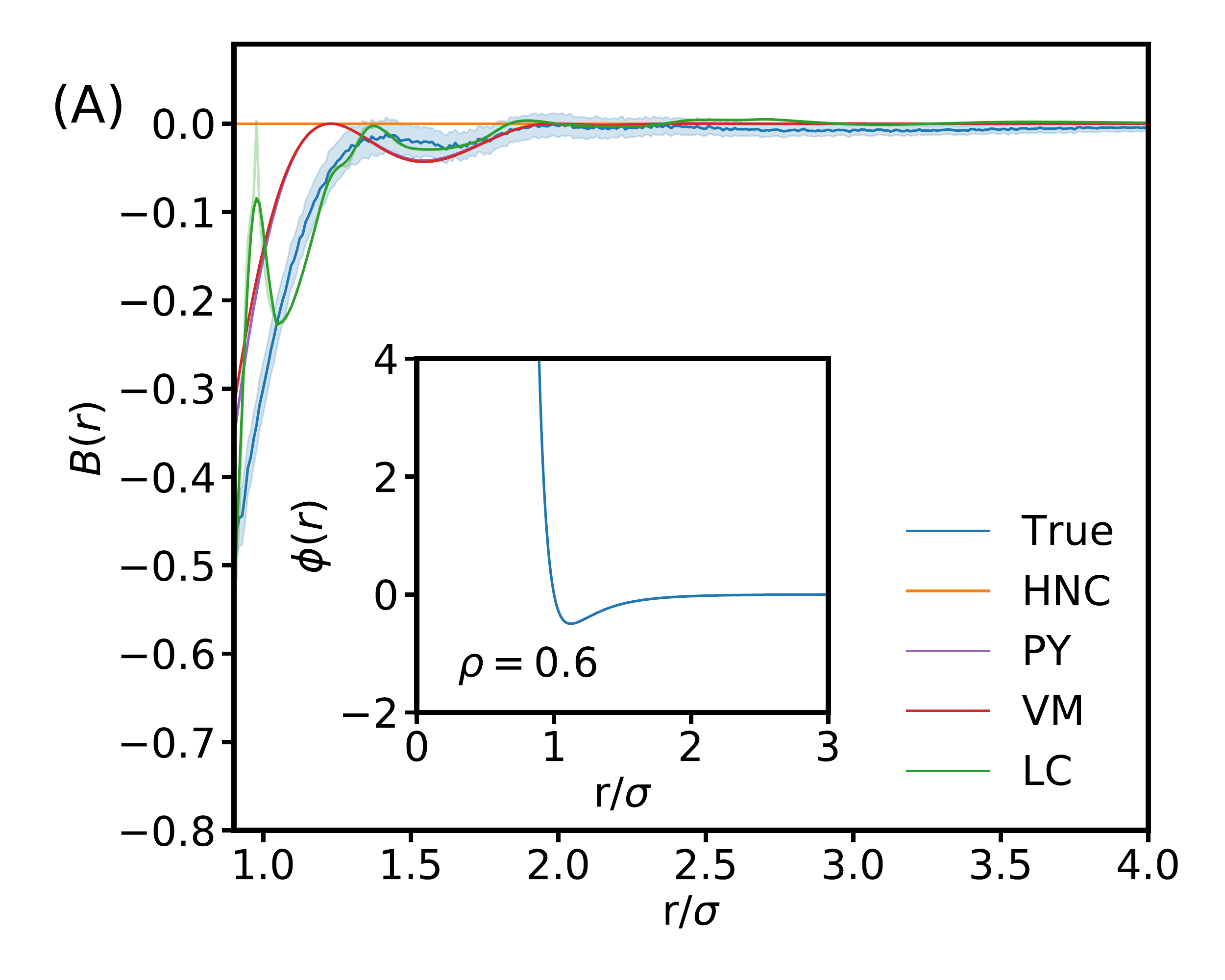}
    \includegraphics[width=0.49\textwidth]{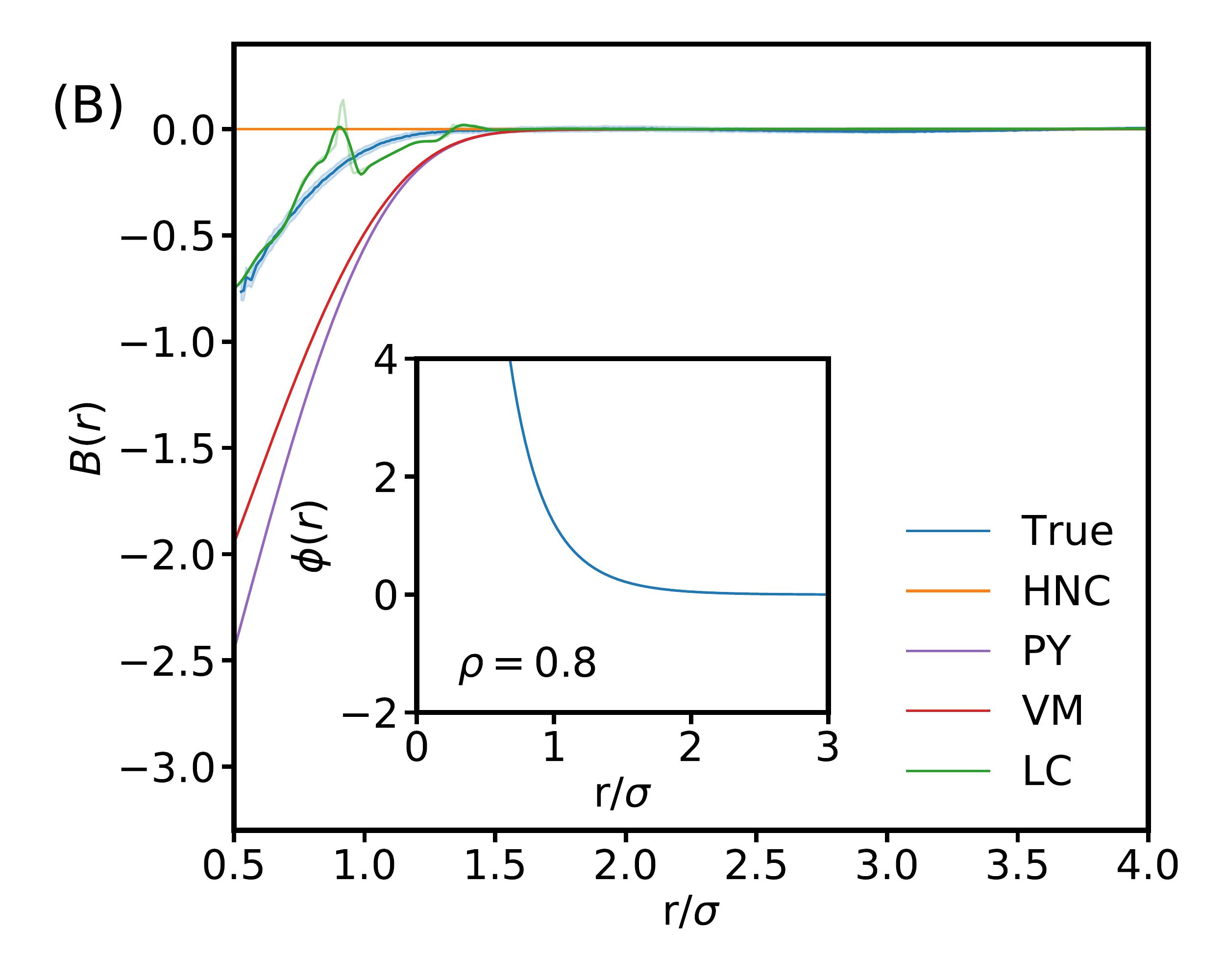}
    \includegraphics[width=0.49\textwidth]{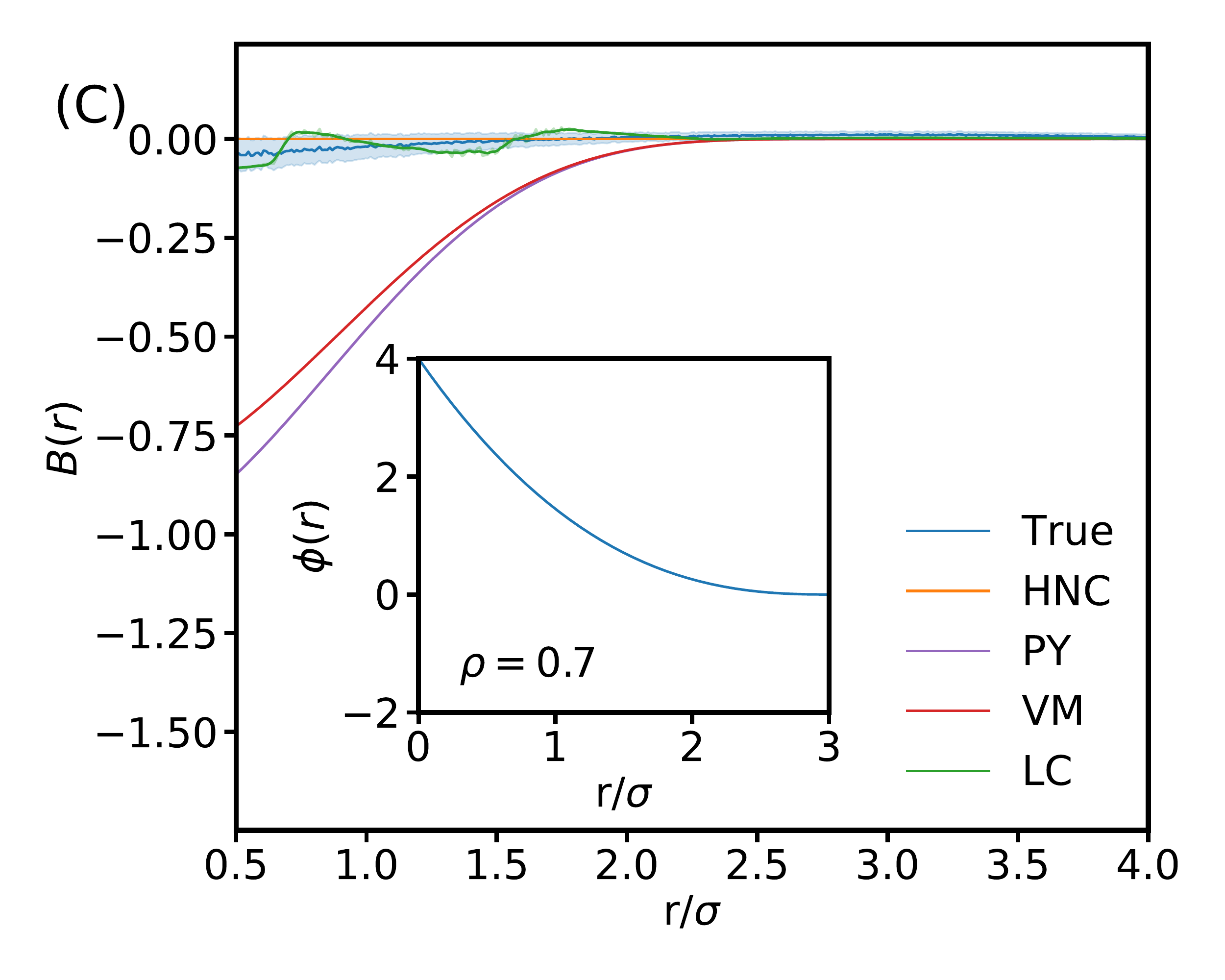}
    \includegraphics[width=0.49\textwidth]{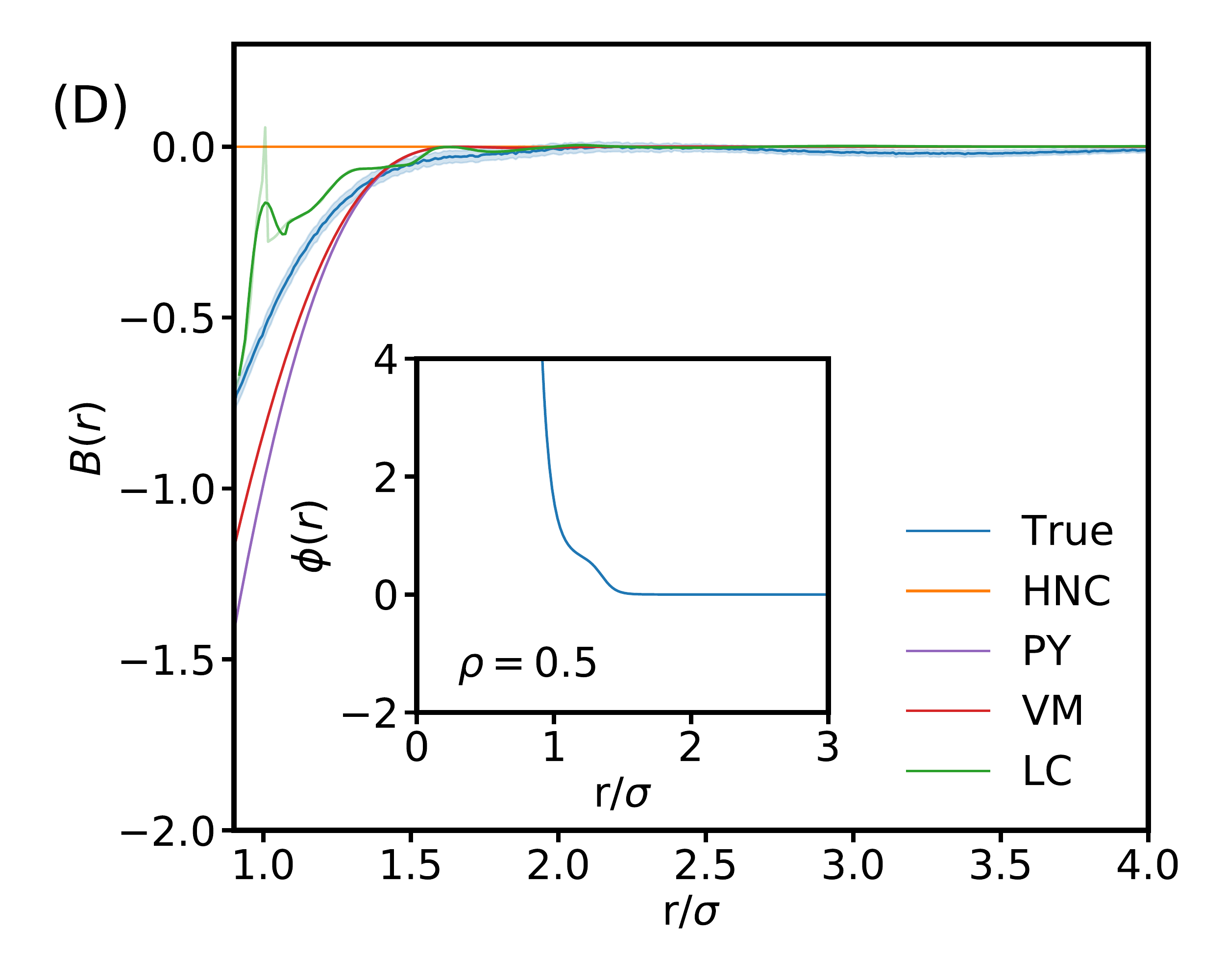}

    \caption{\textbf{Comparisions of learnt and analytical bridge function approximations against the bridge function of known potentials.} \textcolor{black}{The plots show the True bridge function B(r) in blue with the blue shaded region giving the standard error in the B(r). The smoothed LC prediction is shown in green as well as other common bridge functions. The unsmoothed LC prediction is also plotted with high transparency. On all plots, the inset axes show the relevant interaction potential and density of the system. (A) shows a Lennard-Jones system, (B) shows a Yukawa system, (C) a Hertzian system, and (D) an RSSAW system. In all cases, the learnt closure provides a qualitatively closer approximation to the true measurement. However, around the principal peak in $g(r)$ here $r \simeq \sigma$ we see spurious fluctuations in the learnt closure.}}
    \label{fig:bridge-demo}
\end{figure*}

\section{Results}

To examine the performance of learnt closures, we consider \textcolor{black}{and fit MLP models for }the following combinations of the feature set: (1) \textcolor{black}{G $= B(\gamma ; r)$} - a closure just in terms of $\gamma(r)$ as common for analytical closures in the field, (2) \textcolor{black}{HC $= B(h, c ; r)$} - a closure in terms of $h(r)$ and $c(r)$, (3) \textcolor{black}{HCX $= B(h,c,\chi ; r)$} - a closure including $h(r)$, $c(r)$, and $\chi(r)$, and (4) \textcolor{black}{LC $= B(h,c,\chi,\gamma ';r)$} - a learnt closure taking the full feature set as input. To compare these learnt closures to HNC, VM, and PY we look at how they perform on a randomly sampled test set comprising \textcolor{black}{90/450 (20\%) of the simulation systems} that were withheld when training the closures.

Table \ref{tab:compare} shows that the learnt closure based on just the indirect correlation function, G, has a negative coefficient of determination, $R^2$, implying that it is worse than predicting a constant value (i.e. HNC). However, as the feature set is extended to include additional physically motivated features (Fig. \ref{fig:parity-random}A), the learnt closures offer rapidly improving performance compared to HNC\textcolor{black}{, PY, and VM}. Using the full feature set, LC, leads to a very strong correlation between the learnt closure's predictions and the ground truth with a $R^2$ value of \textcolor{black}{0.693}. Physically, this improved performance is due to the learnt closures capacity to capture \textcolor{black}{more} of the strongly correlated physics in the region around the first co-ordination shell \textcolor{black}{across a broad range of interactions.}

\subsection{Generalisation Performance and University}

Our results show that learnt closures can exhibit greater universality \textcolor{black}{than common analytical closures such as HNC, PY or VM} when trained and tested on a diverse selection of potential systems. However, an interesting question is whether this is true generalisation performance that would extend to out-of-training-distribution regimes. This can be probed by examining the generalisation performance of learnt closures trained on restricted classes of potential. Two \textcolor{black}{LC closure models} were trained using the full feature set for this purpose, one on only hard potentials and a second on only soft potentials (Fig. \ref{fig:parity-random}B). When tested in this manner, the learnt closures are seen to be less accurate in their out-of-training-distribution regimes. This result suggests that it would not be reasonable to apply learnt closures in applications involving qualitatively different potentials (e.g. charged liquids, where the correlation length-scales are much longer) without first extending the training data to also include such systems.

\textcolor{black}{We can also look at how well the model performs on individual systems. For an LC model trained on all the training data Fig. \ref{fig:bridge-demo} shows the true bridge function and common approximations plotted alongside the LC model's predictions for example Lennard-Jones, Yukawa, Hertzian, and RSSAW systems. Whilst the LC model offers qualitatively improved performance for all of these different interactions, despite not having been exposed to training data at the state points being examined, we observe spurious fluctuations around the principal peak in $g(r)$. The presence of these fluctuations suggests that the feature set proposed in this work is insufficient to build a truly universal closure that can completely capture all the possible behaviour seen in this region.} As such, whilst learnt closures can exhibit the greater universality this is only a partial solution to the transferability problem that arises in the construction of better analytic closures.

\begin{figure*}
    \centering
    \includegraphics[width=0.99\textwidth]{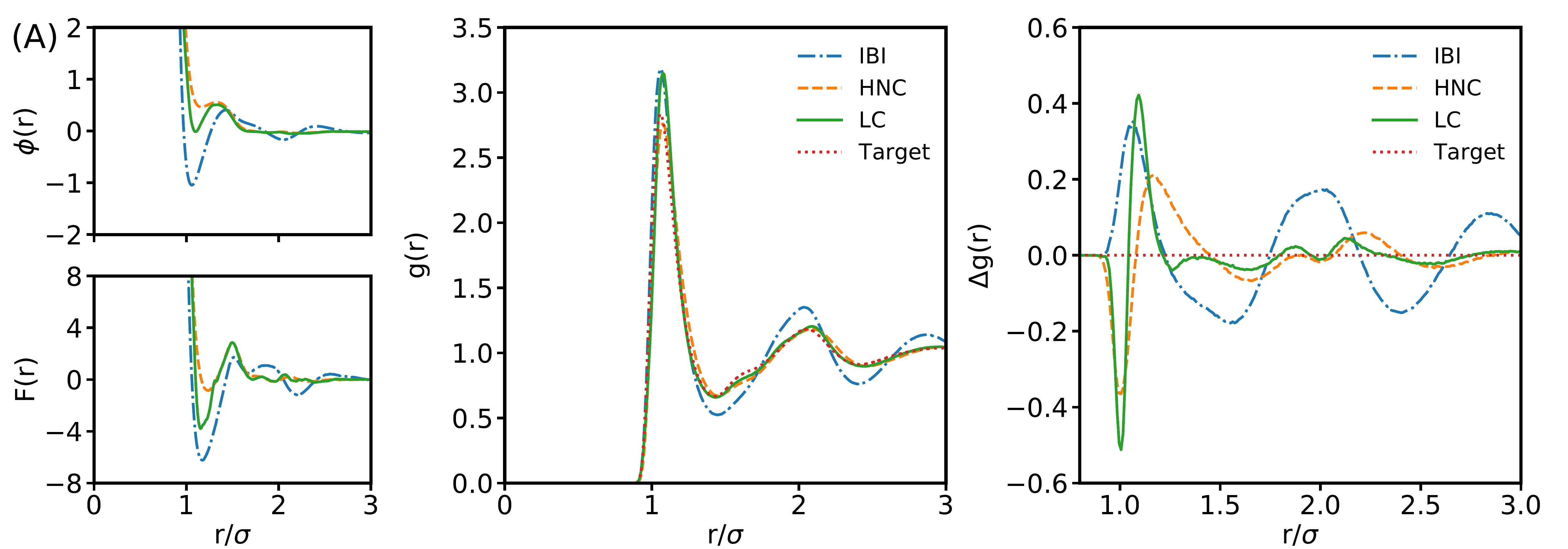}
    \includegraphics[width=0.99\textwidth]{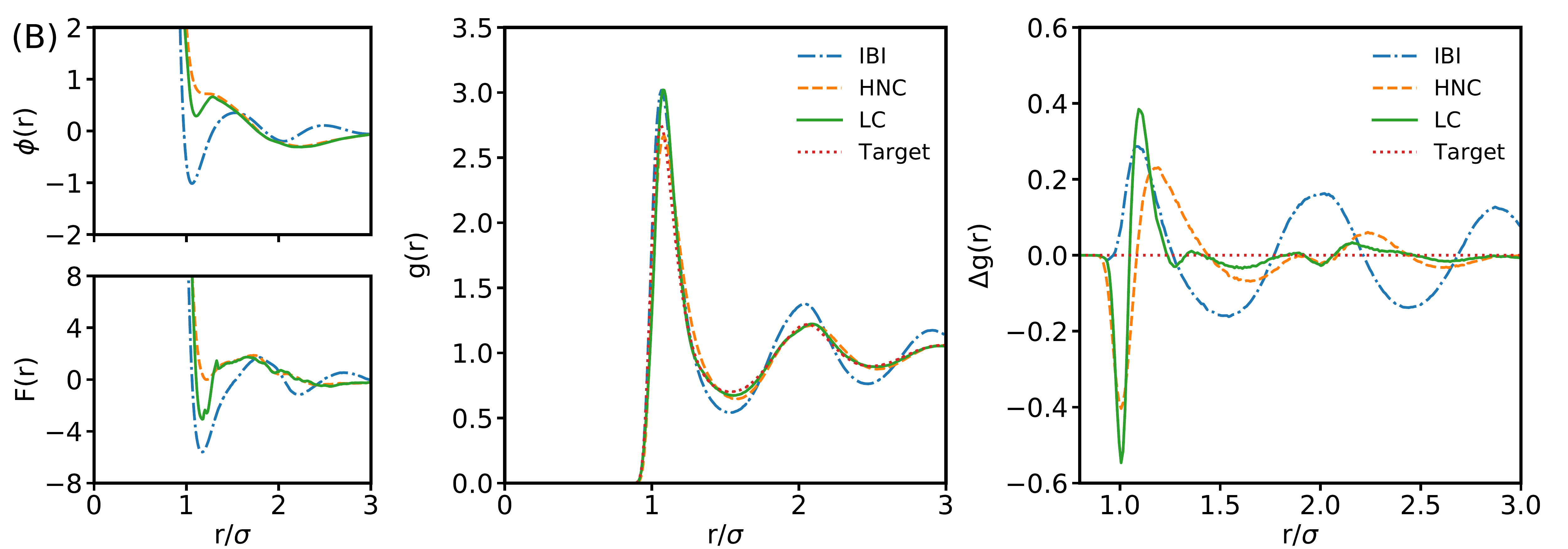}
    \caption{\label{fig:downstream}\textbf{Comparison of learnt closures against IBI and HNC in downstream applications.} \textcolor{black}{(A) shows results for the first toy problem and (B) shows results for the second toy problem. For each system, the small plots on the left-hand side show the initial estimates for the potential $\phi(r)$ and force $F(r)$ using IBI, HNC, and the learnt closure (LC) (same colour scheme as larger plots). The central plots show the resulting $g(r)$ for each of the initialisations as well as the original target distribution. In both examples, IBI and LC result in distributions that overestimate the height of the principal peak while HNC underestimates the peak. The rightmost plots show the differences between the resulting $g(r)$ and the target distribution.}}
\end{figure*}

\subsection{Coarse-graining with machine learnt closures}

To test the potential benefits of using a learnt closure in downstream applications, we consider the prototypical task of coarse-graining the solvent degrees of freedom in a two component solvent-solute mixture. The challenge is determining the effective solute-solute interaction such that the resulting solute-only (one-component) simulation reproduces the solute pair distribution function of the underlying solute-solvent (two-component) system. Such inverse problems are usually solved using iterative methods such as Iterative Boltzmann Inversion (IBI) and multi-state variants thereof \cite{reith2003deriving, moore2014derivation}. Recent work has extended IBI to make use of the Ornstein-Zernike framework in Iterative Ornstein-Zernike Inversion (IOZI) \cite{heinen2018calculating} where the iteration scheme relies on the use of a closure approximation to close the equations. Both processes work by running forward simulations to find the liquid structure that results from a given test potential. The resulting structures are then used to iteratively update the test potential in order to minimise deviations between the observed structure and the desired structure according to the update rule,
\begin{align}\label{eq:update}
    \phi_{n+1}(r)   &= \phi_{n}(r) + k_bT \ln\bigg(\frac{g_n(r)}{g^*(r)}\bigg) 
    + \gamma^*(r) - \gamma_n(r) \nonumber \\
                    &\quad + \hat{B}^*(r) - \hat{B}_n(r) ,
\end{align}
where $\hat{B}(r)$ denotes an estimate of $B(r)$ \textcolor{black}{using a given approximation for the bridge function}. The IBI update rule corresponds to keeping just the first two terms on the right hand-side of this iteration scheme. All of these methods rely on making some initial estimate for the potential. For IOZI the initial estimate is: 
\begin{equation}\label{eq:iozi_init}
    \phi_1^{IOZI}(r) = - k_bT \ln\big(g^*(r)\big) + \gamma^*(r) + \hat{B}^*(r),
\end{equation}
 whilst IBI typically uses: $\phi_1^{IBI}(r) = - k_bT \ln\big(g^*(r)\big)$.

Improved initial estimates have the potential to significantly speed up the convergence of the iteration process \cite{mashayak2018integral}. Indeed, a true closure to the Ornstein-Zernike formalism, one yielding the correct $B^*(r)$, would converge in one iteration. As such, a valid comparison of different approaches can be made purely on how well the initial estimate reproduces a structure that matches the target. 

Our \textcolor{black}{first} toy problem consists of equal quantities of two species of Lennard-Jones particles \textcolor{black}{where one species has twice the radius of the other. The energy and lengthscales for the interactions were $\epsilon_{00} = 1.0, \epsilon_{01} = 1.1$, $\epsilon_{11} = 1.0$, $\sigma_{00} = 1.0, \sigma_{01} = 0.75,$ and $\sigma_{11} = 0.5$. The cutoff distances were $r_{00}^{cut} = 2.5, r_{01}^{cut} = 2.5,$ and $r_{11}^{cut} = 1.5$. Both species had particle densities of $\rho=0.8$.}

\textcolor{black}{The second toy problem consists of a set of Lennard-Jones (LJ) particles and a set of smaller Weeks-Chandler-Andersen (WCA) particles that interact with each other by a Smooth Step (ST) potential. The parameters for all three interactions are $\epsilon^{LJ} = 1.0$, $\sigma^{LJ} = 1$, $r^{LJ}_{cut} = 2.5$, $\epsilon^{WCA} = 1.0$, $\sigma^{WCA} = 0.6$, $d^{ST}=0.7$, $n^{ST}=10$, $\epsilon_h^{ST}=1$, $\epsilon_s^{ST}=-1$, $\kappa^{ST}=2.5$, $\delta^{ST}=1.5$, $\sigma^{ST} = 1$, and $r^{ST}_{cut}=3$. Note that setting $\epsilon_{ST}$ to be negative gives rise to a broad attractive minimum and not a repulsive step. As before both species had particle densities of $\rho=0.8$. For both toy problems the rest of the simulations details were exactly as described in the Supplementary Materials.}

\textcolor{black}{We measured $g^*(r)$ and $S^*(q)$ between the larger species in the first problem and between the particles interacting via the LJ potential in the second. These correlation functions were used to calculate $c^*(r)$ and in turn estimate $\hat{B}^*(r)$ using an LC model from which $\phi_{1}^{LC}(r)$ was determined according to \eqref{eq:iozi_init}}. As the bridge functions are estimated in a point-wise manner the resulting potential contains some high-frequency noise that affects its numerical gradient. \textcolor{black}{As the the force is given by $F(r)= -\nabla\phi(r)$ to} obtain well-behaved forces we apply a \textcolor{black}{quadratic} Savitzky-Golay filter \textcolor{black}{with window size 15} to first smooth the estimated potentials before taking their gradients \cite{savitzky1964smoothing}. \textcolor{black}{These interaction potentials and the resulting radial forces are plotted in the leftmost panels of Fig. \ref{fig:downstream}.}

\textcolor{black}{Having obtained different initial estimates for an effective one-species interaction potential using the IBI initialisation and the IOZI initialisation with both HNC and a learnt LC approximation we then simulate, using the same setup, systems of particles with the same density as the species for which the correlation functions were measured interacting via these estimated potentials.} Fig. \ref{fig:downstream} shows the pair distributions resulting from different \textcolor{black}{potentials. Whilst the learnt approximation leads to the height of the principal peak being overestimated in both examples} it does much better at matching the complex step-like structure of the $g^*(r)$ around $1.6\sigma$ \textcolor{black}{in the first example}. To quantify the \textcolor{black}{improvement} we can look at the Wasserstein distance between the target particle density, $r^2 g^*(r)$, and the density that results from the initialisation, $r^2 g_1(r)$. This metric is directly related to the amount by which particles would need to be moved to recover the target density. The resulting distances are; IBI - 0.243, HNC - 0.044 and, LC - 0.037 \textcolor{black}{for the first example and IBI - 0.306, HNC - 0.053 and, LC - 0.038 for the second}. These results show quantitatively benefits of the learnt closure over IBI and HNC.

\section{Conclusions}

In this work, we demonstrate that machine learning is an effective tool for tackling inverse problems in soft matter. We use the physics-derived framework of Ornstein-Zernike theory but employ machine learning to parameterise its closure relationship using physically \textcolor{black}{relevant correlation functions}. Our approach can be used to construct closures that are accurate in regions where traditional analytical approximations tend to fail. We show that learnt closures can predict the bridge function to sufficient accuracy to have meaningful benefits in scientifically interesting downstream applications, such as the coarse-graining of multi-species systems. We envisage that future work will be able to generalise and extend upon the approach adopted here to obtain increased efficacy closures in a wide variety of systems to which the Ornstein-Zernike framework has been applied such as the RISM and Molecular Ornstein-Zernike approaches \cite{blum1972invariant, chandler1972optimized, iet2015review} where HNC is often still the closure relationship of choice \cite{ding2017efficient}.

More broadly, whilst many theoretical frameworks in physical sciences are elegant and exact, the implementation of those frameworks typically requires approximations and fitted functions. This is particularly true in soft matter where timescale and lengthscale challenges necessitate the use of creative approximations. We believe advances abound in approaches that leverage the overall physics framework but employ machine learning to determine those fitting functions directly from data.

\section{Supplementary Material}
\textcolor{black}{The Supplementary Online Material contains the functional forms of the interaction potentials used, details of the general simulation setup using \textit{ESPResSo}, and details about calculation of the direct correlation function.}

\section{Acknowledgements}
REAG and AAL acknowledge the Winton Programme for the Physics of Sustainability for funding. AAL acknowledges support from the Royal Society. Part of this research was performed while REAG was visiting the Institute for Pure and Applied Mathematics (IPAM), which is supported by the National Science Foundation (Grant No. DMS-1440415).

\section{Data Availability}
All the cleaned data produced in this work are shared on \href{https://github.com/CompRhys/ornstein-zernike}{https://github.com/CompRhys/ornstein-zernike}. The raw data is available from the corresponding author upon reasonable request.

\section{Code Availability}
All the simulation code and fitting code needed to reproduce this work are available from \href{https://github.com/CompRhys/ornstein-zernike}{https://github.com/CompRhys/ornstein-zernike}. The modified structure factor code added to the \textit{ESPResSo} \cite{limbach06a, arnold13a} molecular dynamics package for this work is available from \href{https://github.com/CompRhys/espresso}{https://github.com/CompRhys/espresso}.


\bibliographystyle{unsrt}

\bibliography{ref.bib}

\end{document}


\title{Supplementary Material: Machine learnt approximations to the bridge function \\yield improved closures for the Ornstein-Zernike equation}

\author{Rhys E. A. Goodall}
    \affiliation{Cavendish Laboratory, University of Cambridge, Cambridge, UK}
    
\author{Alpha A. Lee}
    \email[Correspondence email address: ]{aal44@cam.ac.uk\\}
    \affiliation{Cavendish Laboratory, University of Cambridge, Cambridge, UK}

\maketitle

\section{Potential Systems}

\subsection{Fast-Diverging Potentials:}
\begin{itemize}
\item Lennard Jones 6-12 (LJ)
\begin{equation}
\phi(r) = 4\epsilon\bigg(\bigg(\frac{\sigma}{r}\bigg)^{12}-\bigg(\frac{\sigma}{r}\bigg)^6\bigg)
\end{equation}
Lennard Jones is the classical potential used when simulating simple systems. It encapsulates two key effects, hard-sphere repulsion and long range Van der Waals attraction.

\textcolor{black}{We study systems with $\epsilon \in \{0.65, 0.6, 0.55, 0.5\}$, and $\sigma=1$.}

\item Morse
\begin{align}
\phi(r)=&\epsilon\big(\exp[-2 \alpha (r - r_{min})] \nonumber \\
&- 2\exp[-\alpha(r - r_{min})]\big)
\end{align}
The Morse potential is qualitatively similar to LJ but allows slightly more freedom to tune the shape of the minimum. It is often used to model the inter-atomic interactions inside diatomic molecules such as $N_2$.

\textcolor{black}{We study systems with $\epsilon \in \{0.65, 0.6, 0.55\}$, $\alpha \in \{5, 9\}$, and $r_{min}=1$.}

\item Generalised Pseudo-Hard-Sphere

The Mei Potential is a generalised form of LJ that offers more freedom to tune the shape of the potential.
\begin{equation}
\phi(r)=
 \psi\left(\frac{\lambda_r}{\lambda_a}\right)^{\psi} \epsilon \bigg(\bigg(\frac{\sigma}{r}\bigg)^{\lambda_r}-\bigg(\frac{\sigma}{r}\bigg)^{\lambda_a}\bigg) 
\end{equation}
where $\psi = \frac{\lambda_r}{\lambda_r-\lambda_a}$.

The Weeks-Chandler-Andersen (WCA) potential \cite{wca1971potential} is defined by truncating and shifting the LJ potential at its minimum, the resulting potential is purely repulsive. We have constructed the equivalent to WCA for the generalised Mei potential allowing us to test purely repulsive behaviour for a variety of exponents. 

\begin{align}
\phi(r)=
&\psi \left(\frac{\lambda_r}{\lambda_a}\right)^{\psi} \epsilon \bigg(\bigg(\frac{\sigma}{r}\bigg)^{\lambda_r}-\bigg(\frac{\sigma}{r}\bigg)^{\lambda_a}\bigg) \\
&- \phi_{Mei}(r_c) 
\end{align}

This form can be used to mimic the discontinuous potential of an idealised hard-sphere system \cite{jover2012pseudo}.

\textcolor{black}{We study systems with $\epsilon \in \{1.0, 0.6\}$, $(\lambda_r, \lambda_a)  \in \big\{(12,6), (50,49)\big\}$, and $\sigma=1$.}

\item DLVO-type potentials

\begin{align}\label{eq:dlvo-1}
\phi(r) = & \epsilon_h \bigg(\frac{\sigma}{r+\delta}\bigg)^{12} - \epsilon_h \bigg(\frac{\sigma}{r+\delta}\bigg)^8  \nonumber \\
&+ \epsilon_w\frac{\sigma \exp(-\kappa ({r+\delta -1})^4)}{r+\delta}
\end{align}

\begin{equation}\label{eq:dlvo-2}
\phi(r) = \epsilon\bigg(\alpha\bigg(\frac{\sigma}{r}\bigg)^{12}-\bigg(\frac{\sigma}{r}\bigg)^8+\bigg(\frac{\sigma}{r}\bigg)^4\bigg)
\end{equation}

These two potentials are invented potentials designed to try and mimic the secondary stable minimum seen in DLVO theory.

\textcolor{black}{For \eqref{eq:dlvo-1} we study a system with $\epsilon_h \in \{3, 5, 7\}$, $\epsilon_w = 1$, $\kappa \in \{10, 20, 30 \}$, $\delta = 0.3$, and $\sigma=1$.}

\textcolor{black}{For \eqref{eq:dlvo-2} we study systems with $\epsilon = 1$, $\alpha  \in \{0.185, 0.2, 0.215, 0.23, 0.245\}$, and $\sigma=1$.}

\end{itemize}

\subsection{Step-Diverging Potentials:}
\begin{itemize}
\item Smooth Step Potential
\begin{equation}
\phi(r)=\epsilon_h\bigg(\frac{\sigma}{r}\bigg)^{12} + \frac{\epsilon_s}{1 + \exp[2\kappa (r - \delta)]}
\end{equation}
Current closure relationships are known to fail for systems with features over multiple length scales making the smooth step a good choice of training system is we want to extend the generalisability of our inferred closure.

\textcolor{black}{We study systems with  $\epsilon_h = 1$,
$\epsilon_s = 1$, $\kappa  \in \{1, 3, 5, 7\}$, $\delta  \in \{1, 1.5, 2\}$, and $\sigma=1$.}

\item Continuous Shouldered Well (CSW)

\begin{align}
\phi(r)=\epsilon_h\bigg(\frac{\sigma}{r}\bigg)^{12} &+ \frac{\epsilon_s}{1 + \exp[2\kappa (r - \delta_s)]} \nonumber \\
&- \epsilon_w \exp\left(-\frac{1}{2}\left(\frac{r-\delta_g}{\chi}\right)^{2}\right)
\end{align}

The CSW model is a core-softened model in the same manner as the smooth step but it has also been shown to recreate physical anomalies seen experimentally in fluids such as water \cite{lukvsivc2017phase}. 

\textcolor{black}{We study systems with $\epsilon_h = 1$, $\epsilon_s = 2$, $\kappa \in \{2.5, 7.5\}$, $\delta_s \in \{1.2, 1.6\}$, $\epsilon_w = 1$, $\delta_g  = 2.0$, $\chi \in \{0.1, 0.2\}$, and $\sigma=1$.}

\item Repulsive Shoulder System Attractive Well (RSSAW)

\begin{align}
\phi(r)= \epsilon_h \bigg(\frac{\sigma}{r}\bigg)^{14} &- \lambda_1 \tanh(k_1[r-\sigma_1]) \nonumber \\
&+ \lambda_2 \tanh(k_2[r-\sigma_2])
\end{align}

The RSSAW model is similar to the CSW model and exhibits the same complex behaviour \cite{fomin2011complex}. Potentials of this form have been reported for colloidal particles and polymer-colloid mixtures making them important for the study of soft matter systems.  

\textcolor{black}{We study systems with $\epsilon_h = 1$,
 $\lambda_1 = 0.5$, $\lambda_2 = 0.3$, $\kappa_1 = 10$, $\kappa_2 = 10$, $\sigma_1 \in \{0.8, 1.15, 1.5\}$, $\sigma_2 \in \{1, 1.35\}$, and $\sigma=1$.}

\end{itemize}

\subsection{Slow-Diverging Potentials:}
\begin{itemize}
\item Soft-Sphere
\begin{equation}
\phi(r)=\epsilon\bigg(\frac{\sigma}{r}\bigg)^n
\end{equation}
The soft-sphere potential is purely repulsive but allows for more interpenetration than other repulsive models.

\textcolor{black}{We study systems with $\epsilon \in \{1, 6\}$, $n \in \{4, 6, 8, 10\}$, and $\sigma=1$.}

\item Yukawa
\begin{equation}
\phi(r) = \frac{\epsilon \exp\big(-\kappa (r-\delta)\big)}{r}
\end{equation}
The Yukawa potential is a screened coulomb potential that is used to represent the effect of charges in ionic solutions.

\textcolor{black}{We study systems with $\epsilon \in \{2, 4, 6\}$, $\kappa \in \{2.5, 3.5\}$, and $\delta=0.8$.}
\end{itemize}

\subsection{Core-Overlapping Potentials:}
\begin{itemize}

\item Hertzian
\begin{equation}
\phi(r)=\epsilon\left(1-\frac{r}{r_c}\right)^{5/2}
\end{equation}
The Hertzian potential effectively describes the interactions between weakly deformable bodies such as globular micelles. Soft-core potentials are qualitatively different from soft-sphere potentials in far as complete overlap is allowed. 

\textcolor{black}{We study systems with $\epsilon \in \{4, 6, 8, 10\}$ and $r_c \in \{2, 3\}$.}

\item Hat
\begin{equation}
\phi(r)=F_{max} \cdot \frac{r-r_c}{\sigma} \cdot \left( \frac{r+r_c}{2r_c} - 1 \right)
\end{equation}
The Hat potential a is standard conservative potential often used in Dissipative Particle Dynamics for simulating coarse grained fluids.

\textcolor{black}{We study systems with $F_{max} \in \{4, 6, 8, 10\}$, $r_c \in \{2, 3\}$, and $\sigma=1$.}

\item Gaussian
\begin{equation}
\phi(r) =\epsilon \exp\left(-\frac{1}{2}\left(\frac{r}{\sigma}\right)^{2}\right)
\end{equation}
Gaussian shaped potentials have been used as reasonable approximations for the effective interaction between the centres of polymer chains (Flory-Krigbaum potential \cite{flory1950gaussian}).

\textcolor{black}{We study systems with $\epsilon \in \{4, 6, 8, 10\}$ and $1/\sigma^2 \in \{1, 1.5, 2\}$.}
\end{itemize}

\section{Simulation Details}

ESPResSo is a highly flexible open-source Molecular Dynamics package designed for the simulation of soft matter systems. The simulation engine is written in C and C++ but is controlled via a Python interface.

Interaction potentials were specified using regularly spaced tabulated values, these are linearly interpolated in the core to evaluate the forces at each time step. 

For each state point investigated the density was specified and the box size was fixed at 20 $\sigma$ where $\sigma$ is the scale parameter of the density which we set to 1. The number of particles under consideration was scaled accordingly. We consider reduced particle densities in the range 0.4-0.8 at increments of 0.1.

The system is integrated using the Velocity Verlet algorithm \cite{verlet1967integration, verlet1982velocity}. The resulting global errors in the velocities and positions are $O(\Delta t^2)$. A time step of $\Delta t = 0.005$ was used in this work. 

A Langevin thermostat is used to control the system temperature. The Langevin thermostat introduces stochastic momentum fluctuations that both regulate the temperature and are necessary to recreate the fluctuations observed in the canonical ensemble (NVT) making it superior to a rescaling thermostat that would suppress such fluctuations. We opt to set the temperature of the thermostat to 1 and adjust energy scales in the parameterisations of the potentials.

To minimise the chances of a quasi-stable solid phase forming particle positions are randomly initialised to ensure a low symmetry starting arrangement. At the start of each simulation run a static energy minimisation is performed via gradient descent without the thermostat to remove any overlaps present between hard-sphere potentials. \textcolor{black}{Warm-up} runs were then carried out with the thermostat to allow the system to equilibrate. Equilibrium was taken to be the point at which the kinetic temperature over a short windowing period is consistent with the reference temperature of the thermostat.

For computational efficiency the potentials used are truncated at $r_{cut}=3$. The potentials have also been adjusted to fix the potential and force at the cut-off. Often this treated with caution as it introduces systematic errors when measuring the thermodynamic properties of a reference system. However, as the structural correlation functions are causally determined by Newton's laws their validity is unaffected by adjusting the potential. Verlet lists are used to efficiently handle the truncation of the potential. A skin length of 0.2 times the cut-off length was chosen in line with common practice \cite{Frenkel2001UMS}.

The radial distribution function and structure factor were measured from the simulation. To get the structure factor in a timely manner we only take measurements along \{100\} type directions within the system such that evaluating $S(q)$ is $O(N)$ in the number of particles. The default approach that evaluates $S(q)$ for every valid grid point scales as $O(N^3)$. In total for each state point 1024 samples were taken at intervals of 16 time steps. The variances were handled using the Flyjberg-Peterson blocking approach \cite{flyvbjerg1989error}.

\section{Reliable calculation of the Direct Correlation function}

\textcolor{black}{Following from the Fourier transform of the Ornstein-Zernike equation the} direct correlation function, $c(r)$, can be evaluated from measurements of the static structure factor, $S(q)$, based of the relationship that:
 
\begin{equation}\label{eq:C_q}
c(r) = iFT\bigg(\frac{1}{\rho} \bigg(1- \frac{1}{S(q)}\bigg)\bigg)
\end{equation}
 
In simulation studies, the most common approach for calculating the $S(q)$ is taking the Fourier transform of the total correlation function.

\begin{align}
\label{eq:struct}
S(q) &= 1 + \rho H(q) \nonumber \\ 
        &= 1 + \frac{4 \pi \rho}{q} \int_{0}^{\infty} h(r) r \sin(qr) \ dr
\end{align}
\textcolor{black}{Where $H(q)$ is the Fourier transform of $h(r)$.} However, the minimum image convention means $h(r)$ can only be measured up to half the box length. This limit truncates the domain of the Fourier transform leading to finite size effects. Of the possible finite size effects incurred by truncation the most significant is that the apparent $S(q)$ is not guaranteed to be non-negative in the limit $q \to 0$ \cite{frenkel2013dark}. These artefacts result in large-amplitude long-wavelength fluctuations in $c(r)$ that are inconsistent with the limiting behaviour $\lim_{r \to \infty} c(r) \simeq -\beta\phi(r)$.

In previous work approximate extension schemes \cite{Jolly1976extend} have been used to extend $h(r)$ to infinity to avoid such issues. However, such extension schemes rely on the use of a pre-determined closure. The other approach is to calculate $S(q)$ directly from the Fourier transform of the density. However, this approach is computationally much more expensive and is subject to significant noise in the expected value of $S(q)$ for high wave vectors resulting in short-wavelength fluctuations in $c(r)$. 

\begin{figure}
\centering
\includegraphics[width=3in]{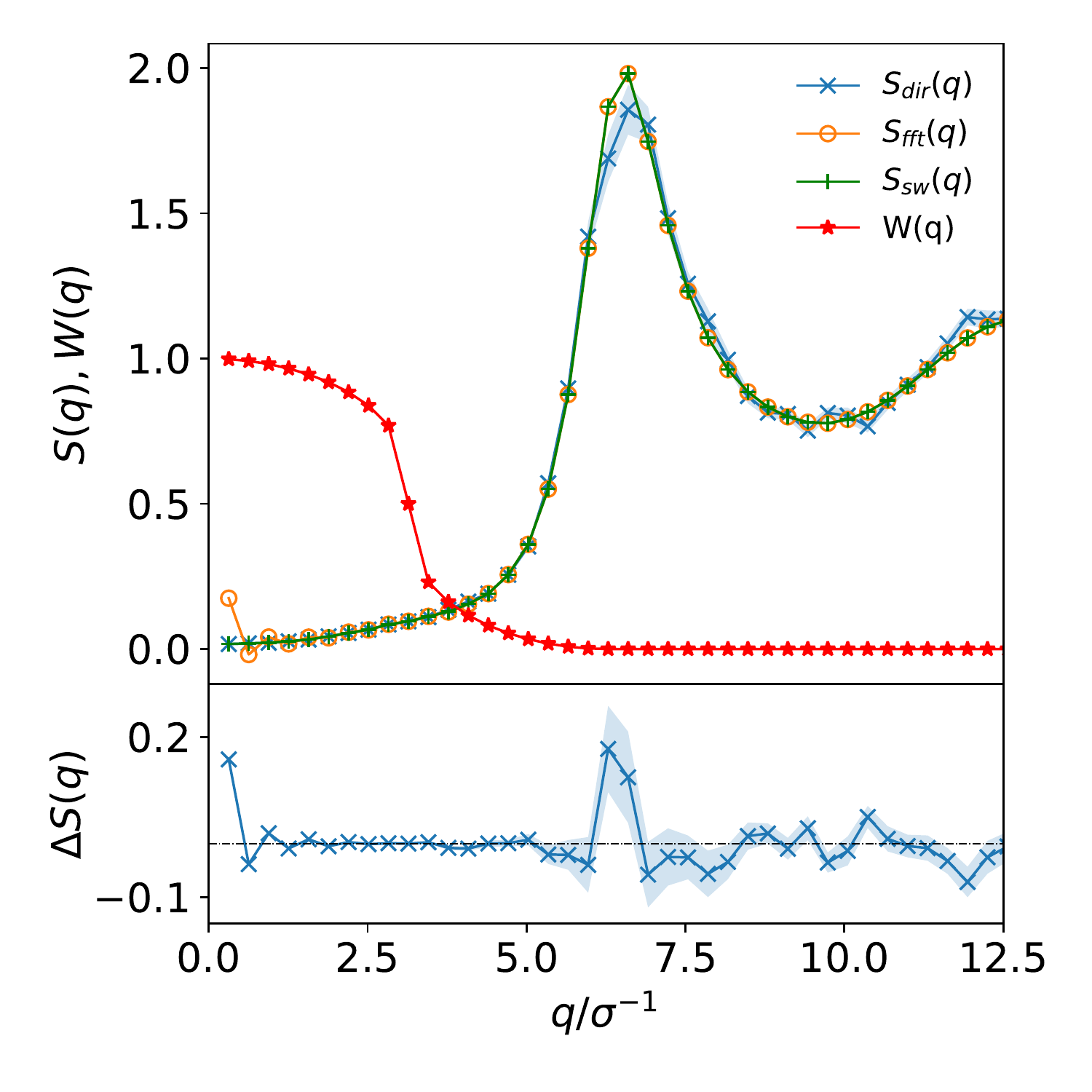}
\caption{\label{fig:hybrid}\textbf{Hybrid scheme to calculate the structure factor.} The upper section shows $S(q)$ as determined directly and from the Fourier transform and shows how the two results are joined together using the switching function W(q). The lower section is a detail of the difference between the direct and Fourier methods for calculating $S(q)$. The figure clearly shows the large oscillations in the low $q$ limit that our method helps to tackle but example also shows the deviation around the principal peak in $S(q)$ which our approach fails to address in some cases.} 
\end{figure}

In this work to avoid both of these limitations, we opt for a Poisson re-summation inspired approach where we evaluate $S(q)$ directly for small wave vectors and from the Fourier transform of $h(r)$ for large wave vectors. A smooth cosine switching function is used to blend between the two regimes. This approach ensures we get the correct limiting behaviour in the $q \to 0$ limit for high-density systems suppressing the non-physical artefacts that would otherwise observed in $c(r)$ if a naive approach was adopted. However, this approach does come with its own limitations as we need to define the switching point heuristically. To reduce potential artefacts from the switching operation we place the transition point in the region before the principal peak where the best agreement is observed between the two methods of calculating $S(q)$. However, for several systems (e.g. Fig. \ref{fig:hybrid}) we observed deviations between the direct and Fourier transform results around the principal peak which in turn can lead to comparatively small but still undesirable intermediate wavelength fluctuations in $c(r)$.

\bibliographystyle{unsrt}
\bibliography{ref}